\documentstyle[12pt,epsf,cite]{article}
\renewcommand{\baselinestretch}{1.20}
\begin{document}
hep-ph/0006212
\begin{flushright}
  June, 2000 \ \ \\
  OU-HEP-352 \ \\
\end{flushright}
\vspace{0mm}
\begin{center}
\large{Polarized Structure Functions $g_2 (x)$ in the
Chiral Quark Soliton Model}
\end{center}
\vspace{0mm}
\begin{center}
M.~Wakamatsu\footnote{Email \ : \ wakamatu@miho.rcnp.osaka-u.ac.jp}
\end{center}
\vspace{-4mm}
\begin{center}
Department of Physics, Faculty of Science, \\
Osaka University, \\
Toyonaka, Osaka 560-0043, JAPAN
\end{center}

\vspace{4mm}

\vspace{6mm}
\begin{center}
\small{{\bf Abstract}}
\end{center}
\vspace{-2mm}
\begin{center}
\begin{minipage}{15.5cm}
\renewcommand{\baselinestretch}{1.0}
\small
\ \ \ The spin-dependent structure functions $g_2 (x, Q^2)$ are
investigated within the framework of the chiral quark soliton
model. It turns out that the twist-3 part of $g_2 (x, Q^2)$ gives
nonnegligible contributions to the total distributions at
the energy scale of $Q^2 = 5 \,\mbox{GeV}^2$ but mainly
in the smaller $x$ region only, so that the corresponding third
moments $\int_0^1 \,x^2 \,{\bar{g}}_2 (x, Q^2) \,dx$ are pretty
small for both of the proton and neutron in conformity with
the recent E155 data.
\normalsize
\end{minipage}
\end{center}
\renewcommand{\baselinestretch}{2.0}

\vspace{8mm}
\ \ \ In our opinion, the unexpectedly small quark spin
fraction of the nucleon indicated by the EMC
experiment \cite{EMC88} and the light flavor sea-quark
asymmetry revealed by the NMC measurement \cite{NMC91}
are two remarkable discoveries in the
field of nucleon structure function physics.
Undoubtedly, both are manifestation of
nonperturbative QCD dynamics imbedded in the physics
of high-energy deep-inelastic scatterings. An outstanding
feature of the chiral quark soliton model (CQSM) is that
it can explain both of these observations without recourse to
any fine-tuning. (This potentiality of the CQSM was
already noticed in \cite{WY91,WAKAM92}.)
In fact, we have already shown that it
reproduces all the qualitatively noticeable features of the
recent high-energy measurements, including the NMC
data for $F_2^p (x) - F_2^n (x)$,
$F_2^n (x) / F_2^p (x)$ \cite{NMC91}, the Hermes and E866
data for $\bar{d} (x) - \bar{u} (x)$ \cite{HERMES,E866},
the EMC and SMC data for $g_1^p (x)$, $g_1^n (x)$ and
$g_1^d (x)$ \cite{E143,E154,E155,SMC99},
in no need of adjustable parameters except for
the starting energy scale of the renormalization-group
evolution equation \cite{WK99,WW00A,WW00B}.

Encouraged by this success, we now push on with our analyses
to the twist-3 parton distributions in the nucleon.
It is well known that, altogether, there are three twist-3
distribution functions -- chiral odd, $e(x, Q^2)$ and
$h_L (x, Q^2)$, and chiral-even,
$g_2 (x, Q^2)$. They are generally believed to provide us with
valuable information on quark-gluon correlations
in hadrons. In the present investigation, we shall focus
our attention on $g_2 (x, Q^2)$. There already exist
several theoretical investigations of the polarized structure
functions $g_2 (x, Q^2)$. The most of those are based on
various modifications of the MIT bag model as well as its
original version \cite{JJ91,STRAT93,JU94,SHT95,SONG96}.
There also exists an investigation based on the
Nambu-Jona-Lasinio soliton model \cite{WGR97}, which is
essentially equivalent to the CQSM. Confining to the lower
moments of $g_2 (x, Q^2)$, also available are theoretical
predictions based on the QCD sum rule \cite{STEIN95,BBK90}
as well as the quenched lattice QCD \cite{GOCKEL96}, etc.

Although based on essentially the same model as \cite{WGR97},
the present investigation goes far beyond the previous one
in many respects. First, the polarization effects of the
negative-energy Dirac-sea quarks in the hedgehog mean-field
are fully taken into account. This is very important for
offering any reliable predictions for antiquark distributions.
Secondly, we also include
the novel $1 / N_c$ correction (or the first order rotational
correction in the collective angular velocity $\Omega$) to
the isovector distribution functions.
Without inclusion of it, some fundamental isovector observables
like the nucleon isovector axial coupling constant would be
largely underestimated, thereby being led to the so-called
``$g_A$ problem'' in the hedgehog soliton
model \cite{WW93,Wakam96}.
Thirdly, the nonlocality effects (in time) inherent in the
theoretical definition of parton distributions are treated
in a consistent way \cite{WK99,PPGWW99}.

We start with the following definition of the distribution
functions :
\begin{eqnarray}
 g_1^{(I=0/I=1)} (x) &=& \frac{1}{2 M} \,
 \int_{- \infty}^{\infty} \,\frac{d \lambda}{2 \pi} \,
 e^{\,i \lambda \,x} \,\langle PS \,\vert \,\psi^\dagger (0) \,
 ( 1 + \gamma^0 \,\gamma^3 ) \,\gamma_5 \,
 \left\{ \begin{array}{c} 1 \\ \tau_3 \end{array} \right\} \,
 \psi (\lambda \,n) \,\vert \,PS \rangle \, , \\
 g_T^{(I=0/I=1)} (x) &=& \frac{1}{2 M} \,
 \int_{- \infty}^{\infty} \,\frac{d \lambda}{2 \pi} \,
 e^{\,i \lambda \,x} \,\langle P S_{\bot} \,\vert \,
 \psi^\dagger (0) \, \gamma^0 \,\gamma_{\bot} \,\gamma_5 \,
 \left\{ \begin{array}{c} 1 \\ \tau_3 \end{array} \right\} \,
 \psi (\lambda \,n) \,\vert \,P S_{\bot} \rangle \, ,
\end{eqnarray}
which is just standard except for the isospin dependence
explained below. Here the parts containing $1$ and $\tau_3$
respectively give isoscalar ($I=0$) and isovector ($I=1$)
combinations of the relevant distributions. They are
normalized such that the corresponding quark distributions
$g_T^{(q)} (x)$ with $q = u, d$ are given as
\begin{equation}
 g_T^{(u/d)} (x) \ = \ \frac{1}{2} \,\,[ \,
 g_T^{(I=0)} (x) \pm g_T^{(I=1)} (x) \,] \hspace{15mm}
 (0 < x < 1),
\end{equation}
and similarly for $g_1 (x)$. The distribution functions (1) and (2)
are formally defined in the region $-1 < x < 1$. The functions
with negative $x$ are to be interpreted as giving antiquark
distributions $g_T^{(q)} (x)$ with $q = \bar{u}, \bar{d}$
according to the rule :
\begin{equation}
 g_T^{(\bar{u}/\bar{d})} (x) \ = \ \frac{1}{2} \,\,
 [ \,g_T^{(I=0)} (-x) \pm g_T^{(I=1)} (-x) \,] \hspace{15mm}
 (0 < x < 1) \, ,
\end{equation}
and similarly for $g_1 (x)$.
The corresponding structure functions for the proton and the
neutron at the model energy scale are then constructed as
\begin{equation}
 g_T^{(p/n)} (x) \ = \ \frac{5}{36} \,[ \,
 g_T^{(I=0)} (x) + g_T^{(I=0)} (-x) \,] \ \pm \ 
 \frac{1}{18} \,[ \,
 g_T^{(I=1)} (x) + g_T^{(I=1)} (-x) \,] \, .
\end{equation}
The twist-2 distribution functions
$g_1 (x)$ were already evaluated in \cite{WK99,WW00A}.
The distribution functions
$g_T^{(I=0)} (x)$ and $g_T^{(I=1)} (x)$ can be evaluated within
the same theoretical framework.
Skipping the detail, here we only recall the fact that the isoscalar
and isovector parts have totally different dependences on the
collective angular velocity $\Omega$ of the rotating hedgehog mean field,
which itself scales as $1 / N_c$ \cite{WK99} :
\begin{eqnarray}
 g_T^{(I=0)} (x) &\sim& N_c \,O (\Omega^1) \ \sim \ O (N_c^0) \, ,\\
 g_T^{(I=0)} (x) &\sim& N_c \,[ \,O (\Omega^0) \ + \ O (\Omega^1) \,]
 \ \sim \ O (N_c^1) + O (N_c^0) \, .
\end{eqnarray}

\vspace{3mm}
\begin{figure}[htb] \centering
\epsfxsize=15.0cm 
\epsfbox{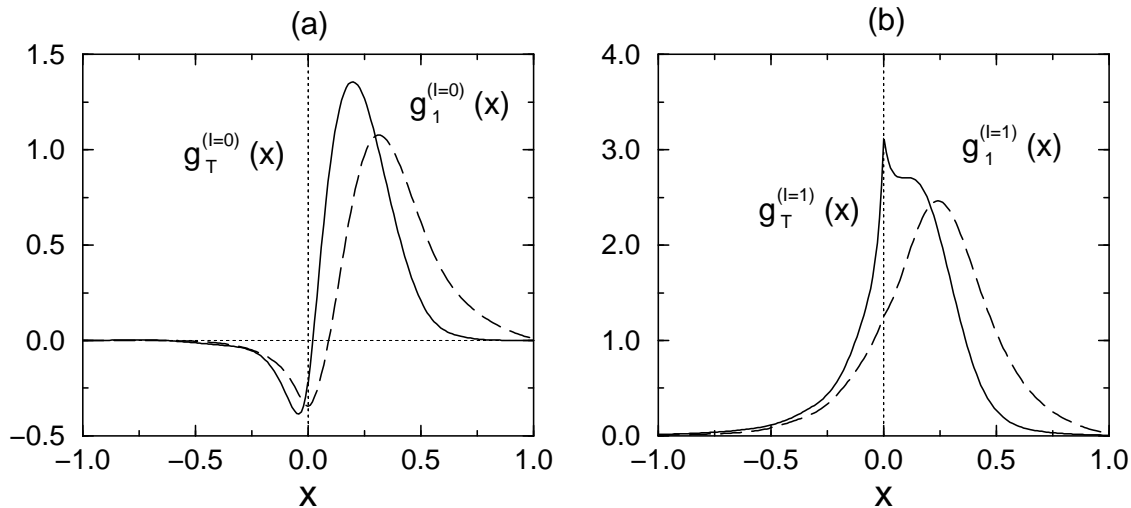}
\renewcommand{\baselinestretch}{1.00}
\caption{(a) The isoscalar parts of the distribution functions
$g_1 (x)$ and $g_T (x)$. (b) The isovector parts of $g_1 (x)$ and
$g_T (x)$.}
\renewcommand{\baselinestretch}{1.20}
\label{fig:fig1}
\end{figure}
\vspace{3mm}

Shown in Fig.1 are the results for $g_T^{(I=0)} (x)$ and
$g_T^{(I=1)} (x)$ in comparison with the twist-2 distributions
$g_1^{(I=0)} (x)$ and $g_1^{(I=1)} (x)$. We first point out that
the isoscalar and isovector distributions have totally dissimilar
shapes reflecting quite different $N_c$-dependence given
in (6) and (7).
Next, comparing $g_T^{(I=0)} (x)$ and $g_1^{(I=0)} (x)$, one
finds that the $g_T^{(I=0)} (x)$ has a peak at smaller value of
$x$ than $g_1^{(I=0)} (x)$ and damps faster as $x$ increases.
The same tendency is also observed for the isovector distributions,
but in this case the concentration of the distribution
$g_T^{(I=1)} (x)$ into the smaller $x$ region is even more profound.
This is due to more significant effect of vacuum polarization,
which is peaked around $x \simeq 0$.
It means that the ``valence-quark-only'' approximation adopted
in \cite{WGR97} cannot be justified at least for $g_T^{(I=1)} (x)$ and
to a lesser extent also for $g_1^{(I=1)} (x)$. Note, however, that
the vacuum polarization contributions, which are dominant in the
smaller $x$ region, are partially canceled in the corresponding
spin structure function $g_2^{(I=1)} (x)$ defined as a difference
of $g_T^{(I=1)} (x)$ and $g_1^{(I=1)} (x)$.
In any case, we emphasize
the following. The CQSM gives fairly different predictions for the
shapes of $g_1 (x)$ and $g_T (x)$. Moreover, the shapes of both
distributions are strongly dependent on the isospin (or more
generally flavor) combinations.
Furthermore, both distributions $g_1^{(I=1)} (x)$
and $g_T^{(I=1)} (x)$ have large support in the negative $x$ region,
which implies sizable flavor asymmetry of the spin-dependent
sea-quark (antiquark) distributions \cite{WW00A,DGPPWW98}.

To compare these predictions of the CQSM with the
existing high-energy data for $g_2^p (x,Q^2)$ and
$g_2^d (x,Q^2)$, we must take account of the scale dependence of
the distribution functions. We have done it as follows.
Remember first that the spin structure function $g_2 (x,Q^2)$
is defined as a difference of $g_T (x,Q^2)$ and
$g_1 (x,Q^2)$, i.e. $g_2 (x,Q^2) = g_T (x,Q^2) - g_1 (x,Q^2)$.
(In the following discussion on the scale dependence, we omit
the isospin indices for the structure functions, for notational
simplicity.)
The Burkhardt-Cottingham sum rule \cite{BC70} holds exactly, i.e.
\begin{equation}
 \int_0^1 \,g_2 (x,Q^2) \,dx \ = \ 0 \, ,
\end{equation}
if the charges (or the first moments) of $g_1 (x,Q^2)$
and $g_T (x,Q^2)$ are equal, which in turn follows from the
rotational invariance of the whole theoretical scheme.
This property is automatically satisfied in usual low
energy models like the MIT bag model \cite{JJ91,STRAT93,
JU94,SHT95,SONG96} or the CQSM \cite{WGR97}.
The $g_2 (x,Q^2)$ is further decomposed into the twist-2
(Wandzura-Wilczek) part and the genuine twist-3 part
as \cite{WW77}
\begin{eqnarray}
 g_2^{WW} (x,Q^2) &\equiv& - \,g_1 (x,Q^2) \ + \ 
 \int_x^1 \,\frac{dy}{y} \,\,g_1 (y,Q^2) \,  \\
 \bar{g}_2 (x,Q^2) &\equiv& g_2 (x,Q^2) \ - \ 
 g_2^{WW} (x,Q^2) \, .
\end{eqnarray}

For the QCD evolution of the structure function $g_1 (x,Q^2)$
and twist-2 piece of $g_2 (x,Q^2)$, the ordinary
Dokshitzer-Gribov-Lipatov-Altarelli-Parisi (DGLAP) equation
can be used. Here, we use the leading-order Fortran code
provided by Saga group \cite{HKM98}.
The flavor non-singlet and singlet
channels are treated separately, since the mixing with the
gluon distribution occurs in the latter.
On the other hand, the $Q^2$-evolution of twist-3 distributions
is known to be quite complicated due to mixing with
quark-antiquark-gluon operators, the number of which increases
rapidly with spin or the moment of the distributions.
However, Ali, Braun and Hiller
found that, in the large $N_c$ limit, the $Q^2$-evolution
of the chiral-even twist-3 flavor-nonsinglet distribution
$\bar{g}_2 (x,Q^2)$ is described by simple DGLAP type equation
with slightly different forms for the anomalous dimensions
from the twist-2 distributions \cite{ABH91}.
Accordingly, the moments of
$\bar{g}_2 (x,Q^2)$ obey the following simple equation :
\begin{equation}
 {\cal M}_n [ \bar{g}_2 (Q^2) ] \ = \ L^{\gamma_n^g / b_0} \,
 {\cal M}_n [ \bar{g}_2 (Q_{ini}^2) ] \, ,
\end{equation}
where ${\cal M}_n [ g(Q^2) ] \equiv \int_{0}^1 \,dx \,x^{n - 1}
\,g (x,Q^2)$, $L \equiv \alpha_S (Q^2) / \alpha_S (Q_{ini}^2)$,
$b_0 = \frac{11}{3} \,N_c - \frac{2}{3} \,N_f$, and
\begin{equation}
 \gamma_n^g \ = \ 2 \,N_c \,\left( \,S_{n-1} - \frac{1}{4} + 
 \frac{1}{2 \,n} \,\right) \, ,
\end{equation}
with $S_n = \sum_{j=1}^n \,\frac{1}{j}$.
The $Q^2$-evolution of the corresponding distribution functions
can be handled by the method described in \cite{KK97}.
In principle, the flavor-singlet part of $\bar{g}_2 (x,Q^2)$
mixes with the gluon distribution, and the $Q^2$-evolution of
it is not given by a simple equation as above even in the
large $N_c$ limit.
In the following study, we shall
neglect this mixing effect with gluons in the twist-3
flavor-singlet distributions, for simplicity.
Finally, the total $g_2 (x,Q^2)$ at the desired energy scale is
obtained after combining the twist-2 and twist-3 pieces of
$g_2 (x,Q^2)$, which are evolved separately.

\vspace{3mm}
\begin{figure}[htb] \centering
\epsfxsize=15.0cm 
\epsfbox{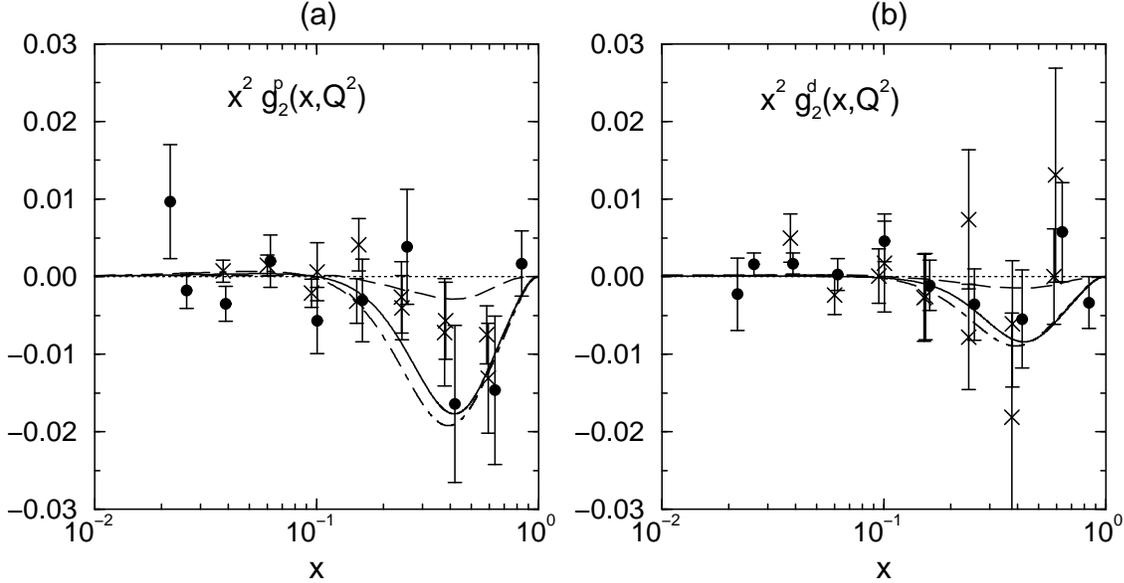}
\renewcommand{\baselinestretch}{1.00}
\caption{The theoretical structure functions 
$x^2 \,g_2^p (x,Q^2)$ and $x^2 \,g_2^d (x,Q^2)$
at $Q^2 = 5 \,\mbox{GeV}^2$ (solid curves) are compared
with the corresponding E143 and E155 data
(respectively shown by crosses and filled
circles) as well as the predictions of Song's center-of-mass
MIT bag model (dashed curves). The twist-2 parts of the
theoretical structure functions (dash-dotted curves) are also
shown for comparison.}
\renewcommand{\baselinestretch}{1.20}
\label{fig:fig2}
\end{figure}
\vspace{3mm}

We show in Fig.2 the theoretical structure functions $g_2 (x,Q^2)$
at $Q^2 = 5 \,\mbox{GeV}^2$ for the proton (a) and the deuteron (b)
in comparison with the corresponding experimental data.
Here, the crosses and the filled circles respectively stand for
the E143 \cite{E143G2} and E155 data \cite{E155G2}.
The final theoretical predictions for
$g_2 (x,Q^2)$ are represented by the solid curves, while
their twist-2 part $g_2^{WW} (x,Q^2)$ are shown by the dash-dotted
ones. The predictions of the center-of-mass MIT bag model by
Song are also shown for the sake of comparison \cite{SONG96}.
We point out that the predictions of the CQSM for both of
$g_2^p (x,Q^2)$ and $g_2^d (x,Q^2)$ are relatively close to those
of another version of MIT bag model given by Strattmann shown
in \cite{E143G2} and sizably larger than those of Song's results.
We also find that,
according to the predictions of the CQSM, the differences between
the full $g_2 (x,Q^2)$ and their twist-2 parts are relatively small
except for the smaller $x$ region, although it is not clear from
Fig.2 in which $x^2$ times $g_2 (x,Q^2)$ are plotted.
This tendency is also close to Stratmann's results rather
than Song's results. 

\vspace{3mm}
\begin{figure}[htb] \centering
\epsfxsize=15.0cm 
\epsfbox{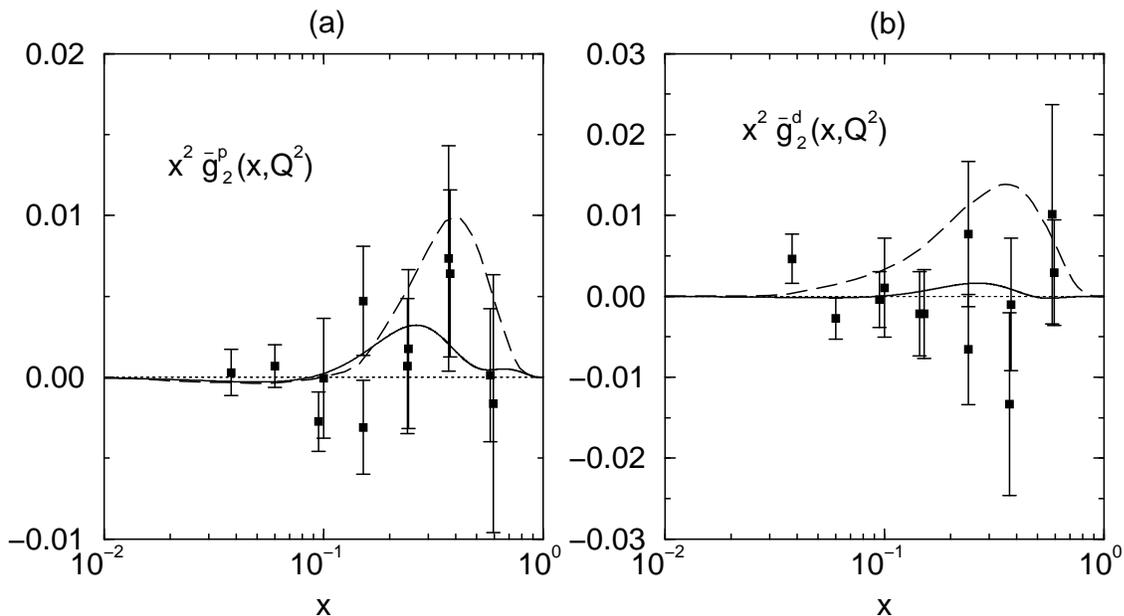}
\renewcommand{\baselinestretch}{1.00}
\caption{The twist-3 parts of the structure functions,
$x^2 \,\bar{g}_2^p (x,Q^2)$ and $x^2 \,\bar{g}_2^d (x,Q^2)$
are compared with Song's predictions as well as the
corresponding experimental data from his paper.}
\renewcommand{\baselinestretch}{1.20}
\label{fig:fig3}
\end{figure}
\vspace{3mm}

To see it more clearly, we show in Fig.3 our
results for the twist-3 part of $g_2^p (x,Q^2)$ and $g_2^d (x,Q^2)$
in comparison with Song's predictions. Here, the solid curves are the
predictions of the CQSM, while the dashed curves are those of the
center-of-mass bag model by Song. The experimental data in this
figure are from \cite{SONG96}.
One clearly sees that the twist-3 parts of
$g_2 (x,Q^2)$ are much smaller in the CQSM than in Song's bag model
calculation. Because of the large uncertainties of the available
experimental data, it is difficult to say  at the present moment
which theoretical prediction is favored.
Nonetheless, small twist-3 contributions to the spin structure
functions $g_2 (x,Q^2)$ appears to be favored by the recent E155
analysis of the twist-3 matrix element :
\begin{eqnarray}
 d_2 (Q^2) \ = \ 3 \,\int_0^1 \,x^2 \,\bar{g}_2 (x,Q^2) \,dx \ = \ 
 2 \,\int_0^1 \, x^2 \,[ \,g_1 (x,Q^2) \ + \ \frac{3}{2} \,
 g_2 (x,Q^2) \,] \,dx \, .
\end{eqnarray}
(Also interesting to notice here would be the fact that the
instanton-liquid model of the QCD vacuum offers a qualitative
explanation of the suppression of the twist-3 matrix element
of $g_2 (x,Q^2)$ relative to the twist-2 one \cite{BPW97}.)
We compare in Fig.4 the predictions of various theoretical calculations
with the recent E155 data. As already shown in \cite{E155G2},
the predictions of some models are apparently incompatible
with the E155 analysis, although we must be cautious about
difficulties in obtaining reliable and precise experimental
information for these quantities.
For instance, large and negative $d_2$ for the neutron
predicted by the QCD sum rules \cite{STEIN95,BBK90} apparently
contradicts the E155 data.
Similarly, large and negative $d_2$ for the proton predicted by the
lattice QCD \cite{GOCKEL96} seems incompatible with the E155 data.

\vspace{3mm}
\begin{figure}[htb] \centering
\epsfxsize=15.0cm 
\epsfbox{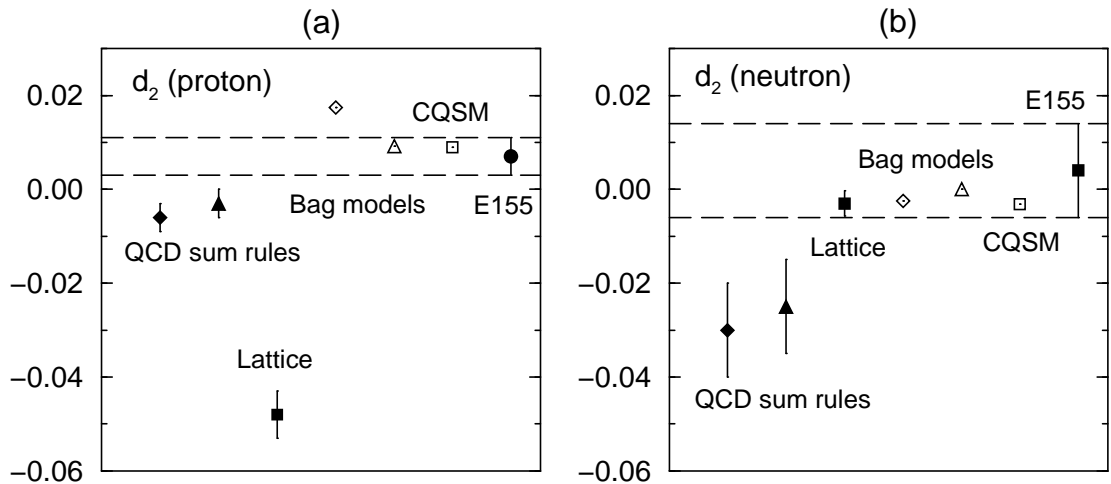}
\renewcommand{\baselinestretch}{1.00}
\caption{The predictions of various theoretical models for the
twist-3 matrix element $d_2$ for the proton (a) and the
neutron (b) are compared with the recent E155 analysis.
Shown theoretical models are from left to right : 
QCD sum rules \cite{STEIN95,BBK90}, lattice
QCD \cite{GOCKEL96}, MIT bag models \cite{SONG96,JU94},
and the CQSM.}
\renewcommand{\baselinestretch}{1.20}
\label{fig:fig4}
\end{figure}
\vspace{3mm}

On the other hand, the predictions of the CQSM as well as those of
the MIT bag models seems consistent with the E155 data at least
qualitatively. We see that the prediction of the naive MIT bag model
for $d_2^p$ is accidentally close to that of the CQSM.
Its prediction $d_2^n = 0$ also seems to lie within the
experimental error bars. Note however that the predictions of
the naive MIT bag model must be taken with care, since the SU(6)
structure of the bag wave function (this is the cause of the result
$d_2^n = 0$ \cite{JU94}) apparently contradicts large and negative
behavior of the twist-2 neutron structure functions $g_1^n (x,Q^2)$
confirmed in several previous experiments \cite{E143,E154,E155,SMC99}.
This shortcoming of the original MIT bag model is remedied in
Song's modified one, in which sizable SU(6) symmetry breaking
effects are incorporated by hand, so that it reproduces both of
$g_1^p (x,Q^2)$ and $g_1^n (x,Q^2)$ \cite{SONG96}.
However, now we sees that the MIT bag model so refined gives a
prediction for $d_2^p$, which is about two times larger than that
of the CQSM and lies outside the errorbars of the E155 data.
More precise experimental data are absolutely awaited for drawing
more decisive conclusion about the twist-3 contributions to the
nucleon spin structure functions, thereby selecting various models
of nucleon internal structure.

To sum up, it has been shown in a series of
paper \cite{WK99,WW00A,WW00B} that the CQSM
reproduces all the qualitatively noticeable
features of the recent high-energy data for the twist-2
structure functions of the proton, the neutron and the deuteron,
with {\it no adjustable parameter} except for the ititial
energy scale of the DGLAP evolution equation.
In the present investigation, we have extended
this parameter-free analyses to the twist-3 spin structure
function $g_2 (x,Q^2)$. The theoretical predictions are
shown to be consistent with the E143 and E155 measurements for
$g_2^p (x,Q^2)$ and $g_2^d (x,Q^2)$ at $Q^2 = 5 \,\mbox{GeV}$,
although the uncertainties of the existing experimental data are
still too large to draw a decisive conclusion. We have also shown
that the CQSM predicts very small twist-3 matrix elements $d_2$ for
the proton and the neutron in conformity with the recent
E155 analysis. The accumulation of more precise experimental
data for $g_2 (x,Q^2)$ as well as $g_1 (x,Q^2)$ is absolutely
necessary for more complete understanding of the nucleon
spin structure.

\vspace{10mm}
\noindent
\begin{large}
{\bf Acknowledgement}
\end{large}
\vspace{3mm}

The author would like to express his sincere thanks to
Y.~Koike for many helpful discussions concerning the scale
dependence of the twist-3 distributions.

%
%

\setlength{\baselineskip}{5mm}

\end{document}